\documentclass[%
 aip,
 amsmath,amssymb,
 reprint,%
]{revtex4-1}

\usepackage{graphicx}
\usepackage{dcolumn}
\usepackage{bm}

\usepackage[utf8]{inputenc}
\usepackage[T1]{fontenc}
\usepackage{mathptmx}
\usepackage{etoolbox}
\usepackage{upgreek}
\usepackage{float}

\makeatletter
\def\@email#1#2{%
 \endgroup
 \patchcmd{\titleblock@produce}
  {\frontmatter@RRAPformat}
  {\frontmatter@RRAPformat{\produce@RRAP{*#1\href{mailto:#2}{#2}}}\frontmatter@RRAPformat}
  {}{}
}%
\makeatother
\begin{document}

\preprint{AIP/123-QED}

\title[Comparison of Nb and Ta Pentoxide Loss Tangents for Superconducting Quantum Devices]{Comparison of Nb and Ta Pentoxide Loss Tangents for Superconducting Quantum Devices}
\author{D.P. Goronzy}
\thanks{These authors contributed equally to this work.}
\affiliation{%
Department of Materials Science and Engineering, Northwestern University, Evanston, IL 60208
}%
\author{W.W. Mah}
\thanks{These authors contributed equally to this work.}
\affiliation{%
Department of Materials Science and Engineering, Northwestern University, Evanston, IL 60208
}%
\author{P.G. Lim}
\affiliation{%
Applied Physics Graduate Program, Northwestern University, Evanston, IL 60208
}%
\author{T. Guess}
\affiliation{%
National Institute for Standards and Technology Boulder, Boulder, CO 80309
}%
\affiliation{%
Department of Electrical, Computer, and Energy Engineering, University of Colorado Boulder, Boulder, CO 80305
}%
\author{S. Majumder}
\affiliation{%
Department of Materials Science and Engineering, Northwestern University, Evanston, IL 60208
}%
\author{D.A. Garcia-Wetten}
\affiliation{%
Department of Materials Science and Engineering, Northwestern University, Evanston, IL 60208
}%
\author{M.J. Walker}
\affiliation{%
Department of Materials Science and Engineering, Northwestern University, Evanston, IL 60208
}%
\author{J. Ramirez}%
\affiliation{%
National Institute for Standards and Technology Boulder, Boulder, CO 80309
}%
\affiliation{%
Department of Physics, University of Colorado Boulder, Boulder, CO 80305
}%
\author{W.-R. Syong}%
\affiliation{%
National Institute for Standards and Technology Boulder, Boulder, CO 80309
}%
\affiliation{%
Department of Physics, University of Colorado Boulder, Boulder, CO 80305
}%
\author{D. Bennett}%
\affiliation{%
National Institute for Standards and Technology Boulder, Boulder, CO 80309
}%
\author{M. Vissers}%
\affiliation{%
National Institute for Standards and Technology Boulder, Boulder, CO 80309
}%
\author{R. dos Reis}
\affiliation{%
Department of Materials Science and Engineering, Northwestern University, Evanston, IL 60208
}%
\affiliation{%
The NUANCE Center, Northwestern University, Evanston, IL 60208
}%
\author{T. Pham}
\affiliation{%
Department of Materials Science and Engineering, Northwestern University, Evanston, IL 60208
}%
\affiliation{%
Department of Materials Science and Engineering, Virginia Polytechnic Institute and State University, Blacksburg, VA 24061
}%
\author{V.P. Dravid}
\affiliation{%
Department of Materials Science and Engineering, Northwestern University, Evanston, IL 60208
}%
\affiliation{%
The NUANCE Center, Northwestern University, Evanston, IL 60208
}%
\affiliation{%
International Institute of Nanotechnology, Northwestern University, Evanston, IL 60208
}%
\author{M.C. Hersam}
\email{m-hersam@northwestern.edu}
\affiliation{%
Department of Materials Science and Engineering, Northwestern University, Evanston, IL 60208
}%
\affiliation{%
Department of Chemistry, Northwestern University, Evanston, IL 60208
}%
\affiliation{%
Department of Electrical and Computer Engineering, Northwestern University, Evanston, IL 60208
}%
\author{M.J. Bedzyk}
\email{bedzyk@northwestern.edu}
\affiliation{%
Department of Materials Science and Engineering, Northwestern University, Evanston, IL 60208
}%
\affiliation{%
Department of Physics and Astronomy, Northwestern University, Evanston, IL 60208
}%
\author{C.R.H. McRae}
\email{coreyrae.mcrae@colorado.edu}
\affiliation{%
National Institute for Standards and Technology Boulder, Boulder, CO 80309
}%
\affiliation{%
Department of Electrical, Computer, and Energy Engineering, University of Colorado Boulder, Boulder, CO 80305
}%
\affiliation{%
Department of Physics, University of Colorado Boulder, Boulder, CO 80305
}%

\date{\today}

\begin{abstract}
Superconducting transmon qubits are commonly made with thin-film Nb wiring, but recent studies have shown increased performance with Ta wiring. In this work, we compare the resonator-induced single photon, millikelvin dielectric loss for pentoxides of Nb ($\mathrm{Nb_2O_5}$) and Ta ($\mathrm{Ta_2O_5}$) in order to further understand limiting losses in qubits. Nb and Ta pentoxides of three thicknesses are deposited via pulsed laser deposition onto identical coplanar waveguide resonators. The two-level system (TLS) loss in $\mathrm{Nb_2O_5}$ is determined to be about 30$\%$ higher than that of $\mathrm{Ta_2O_5}$. This work indicates that qubits with Nb wiring are affected by higher loss arising from the native pentoxide itself, likely in addition to the presence of suboxides, which are largely absent in Ta.
\end{abstract}

\maketitle

\section{\label{sec:intro}Introduction}

Nb is a common choice of material in superconducting transmon qubits,~\cite{Verjauw2021,Anferov2024,Zhao2020} but recent studies have shown that relaxation time $T_1$ increases when either replacing~\cite{Place2021,Wang2021} or simply capping~\cite{Bal2024} Nb with Ta. This indicates that the source of improvement lies within the surface oxides of these two superconductors.

While Nb and Ta native oxides are both fully amorphous,~\cite{Alto2022,McLellan2023} in-depth examinations of Nb and Ta thin films show a significant difference in the presence of suboxides.~\cite{Oh2024} While the Ta native oxide is made up almost exclusively of pentoxide, the Nb native oxide contains an oxygen gradient, with significant amounts of NbO and $\mathrm{NbO_2}$ present near the Nb surface.~\cite{Harrelson2021, Murthy2022, Alto2022} A targeted comparison of the pentoxides allows the distinction between pentoxide- and suboxide-induced losses.

In this work, we compare the loss tangents of coplanar waveguide (CPW) resonators dominated by amorphous pentoxides of Nb and Ta by performing blanket deposition of these materials onto base CPW resonators. We perform in-depth room-temperature structural and chemical characterizations to demonstrate similarity of the deposited pentoxides to native pentoxides and to identify film thickness, homogeneity, and profile. We then measure the two-level system (TLS) and power-independent (PI) loss tangents of these resonators, showing that they increase linearly with pentoxide thickness, as expected for a homogeneous film dominated by bulk loss.

\section{Device Fabrication and Oxide Deposition}

The base device in this study is a DC-sputtered Nb on sapphire CPW resonator of the design as in Ref.~\citenum{Kopas2022}, with a nominal conductor width of 6 $\mu$m and gap of 3 $\mu$m. Details are provided in the supplementary material. No surface treatments are performed prior to pentoxide deposition so that the Nb native oxide is present in all samples in this study (as depicted in Fig.~\ref{fig:matchar}(a)), allowing the base device loss tangent to be identified and separated from the pentoxide loss.

Three thicknesses of $\mathrm{Nb_2O_5}$ and $\mathrm{Ta_2O_5}$ are deposited on these base devices: nominally 3, 15, and 30 nm. The pentoxides are deposited using a room-temperature pulsed laser deposition (PLD) growth with an $\mathrm{O_2}$ partial pressure of 20 mTorr. $\mathrm{Nb_2O_5}$ ($\mathrm{Ta_2O_5}$) is deposited using 390 (701) pulses/nm. The actual thicknesses of each pentoxide film are measured using scanning transmission electron microscopy (STEM, Fig.~\ref{fig:matchar}) and X-ray reflectivity (XRR, Fig.~\ref{fig:XRR_eDensity_XPS}) and discussed in the next section. 

\section{Oxide characterization}

We perform measurements of oxide thickness, conformity, oxidation states, and amorphousness on the deposited films. We show that the deposited oxides are representative of pentoxides found natively on Nb and Ta films.

\begin{figure}
\includegraphics[width=1.05\linewidth]{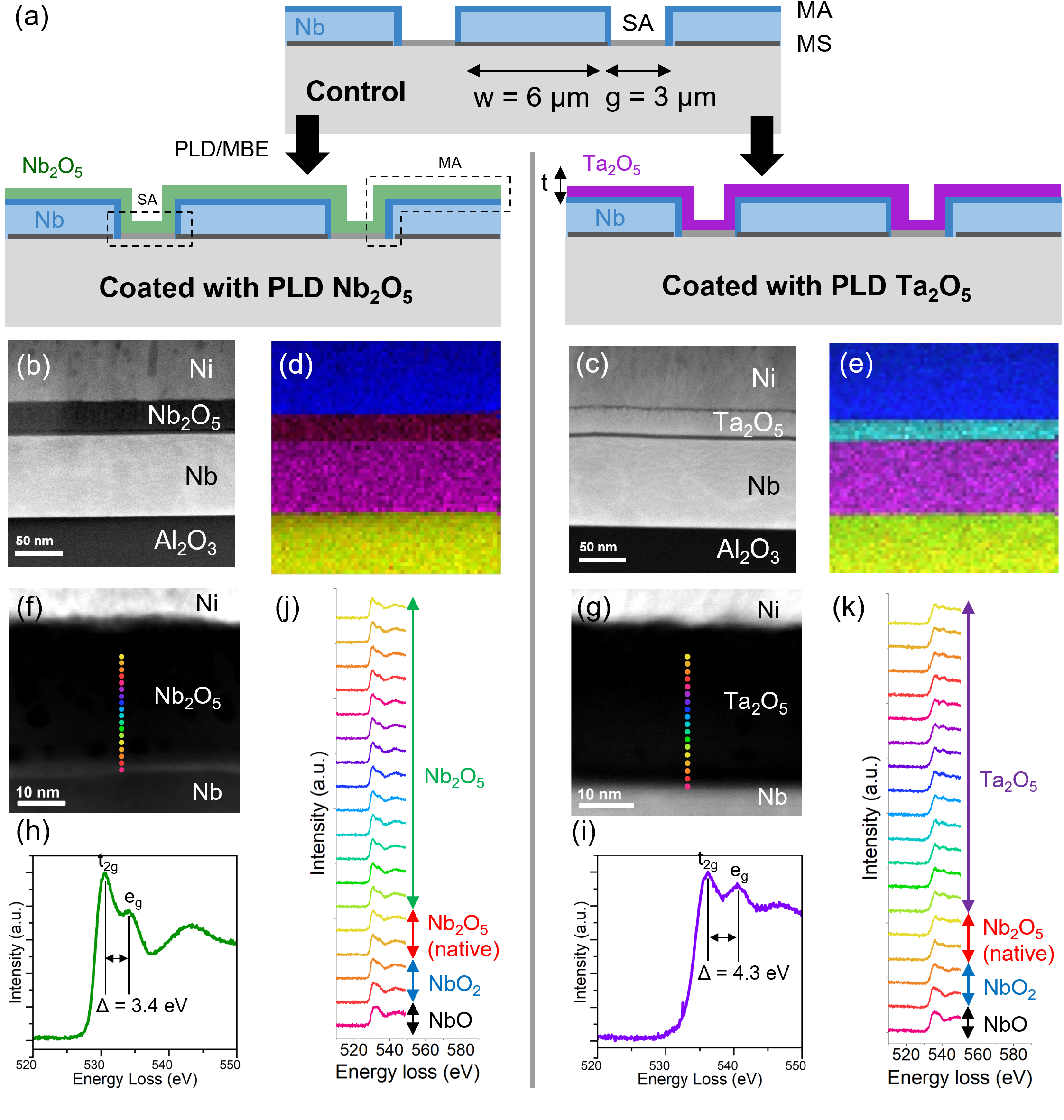}
\caption{\label{fig:matchar}
Pentoxide electron microscopy characterization.
(a) Schematic of control and pentoxide-coated resonator samples. 
(b-c) Representative HAADF-STEM images of the 30 nm (b) Nb\textsubscript{2}O\textsubscript{5}-coated and (c) Ta\textsubscript{2}O\textsubscript{5}-coated Nb films on Al\textsubscript{2}O\textsubscript{3} substrates. 
(d-e) EDS maps corresponding to (b) and (c), respectively, confirm the elemental composition of all layers.
(f-g) High-magnification oxygen K-edge EELS spectrum images of the deposited (f) Nb\textsubscript{2}O\textsubscript{5} and (g) Ta\textsubscript{2}O\textsubscript{5}.
(h-i) ELNES of the oxygen K-edge of the deposited (h) Nb\textsubscript{2}O\textsubscript{5} and (i) Ta\textsubscript{2}O\textsubscript{5}, where the spectrum is summed over the entire deposited pentoxide region. The t\textsubscript{2g}-e\textsubscript{g} crystal field splitting is labeled.
(j-k) EELS line profiles of the oxygen K-edge across the oxide region, including the native Nb oxide and the deposited (j) Nb\textsubscript{2}O\textsubscript{5} and (k) Ta\textsubscript{2}O\textsubscript{5}, corresponding to the color codes in (f) and (g), respectively.
}
\end{figure}

We perform STEM characterization on sample cross-sections prepared by focused ion beam (FIB) to better understand and compare the deposited pentoxides. As expected, Fig.~\ref{fig:matchar}(b) and (c) show $\sim$100 nm of Nb film and $\sim$30  nm of deposited pentoxide, with a thin layer of native Nb oxide between them. We also observe the $\sim$200 nm Ni conductive coating layer, which is deposited prior to FIB sample preparation. This layer is used to protect the oxides from structural and chemical changes induced by ion-beam damage, and to minimize charging and drifting effects. Energy-dispersive X-ray spectroscopy (EDS) measurements (elemental maps in Fig.~\ref{fig:TEM_EDS_SI} superimposed in Fig.~\ref{fig:matchar}(d) and (e)) confirm the chemical makeup of these regions.

To characterize and compare the electronic structure and chemical bonding environment of the deposited pentoxides, we perform electron energy-loss spectroscopy (EELS) at the locations denoted in Fig.~\ref{fig:matchar}(f) and (g). By analyzing the energy-loss near-edge structure (ELNES) of the oxygen K-edge throughout the oxide region (Fig.~\ref{fig:matchar}(h) and (i)), we see that in the deposited pentoxide, there is a very clear energy splitting reminiscent of the transition metal-oxygen bonding environment in $\mathrm{Nb_2O_5}$ and $\mathrm{Ta_2O_5}$, respectively. Moreover, as shown in Fig.~\ref{fig:matchar}(j) and (k), there is a gradual evolution of the ionization edges as we survey the native oxide region near the Nb surface. Specifically, we observe a decrease in the energy splitting between, and changes in the relative intensities, of the two peaks. These changes in the oxygen K-edge near the Nb surface correspond to the different native Nb suboxides present.~\cite{Harrelson2021, Murthy2022, Alto2022, Oh2024} In contrast, the deposited oxides show nearly identical energy loss profiles throughout their thicknesses, confirming that the deposited pentoxides are chemically homogeneous. These spectra are also consistent with those of their respective native pentoxides,~\cite{Bach2006,Oh2024} validating that our targeted comparison of the deposited pentoxides is representative of the native Nb and Ta pentoxides.

Additionally, from EELS, we can extract and compare the energy splitting of the oxygen K-edge doublet peaks for both deposited pentoxides, which we find to be 3.4 eV for $\mathrm{Nb_2O_5}$ and 4.3 eV for $\mathrm{Ta_2O_5}$. These values are lower than those for the crystalline counterparts of these compounds,~\cite{Bach2006,Dickey2006,Tao2011,Harrelson2021,Soriano1993,Kasatikov2019} which is expected for amorphous structures, as decreases in the energy splitting correspond to reduced crystalline character according to ligand-field theory.~\cite{Bach2006,Lafuerza2011,Sassi2017,Frati2020,Harrelson2021} Importantly, the energy splitting of the oxygen K-edge for $\mathrm{Ta_2O_5}$ more closely matches that of crystalline $\mathrm{Ta_2O_5}$ (an 8$\sim$9\% reduction) than does $\mathrm{Nb_2O_5}$ (a 15$\sim$24\% reduction), suggesting that the transition metal-oxygen bonding environment of our amorphous $\mathrm{Ta_2O_5}$ is more similar to crystalline $\mathrm{Ta_2O_5}$ than the PLD deposited amorphous $\mathrm{Nb_2O_5}$ is to crystalline $\mathrm{Nb_2O_5}$, which is in line with previous reports.~\cite{Oh2024}

\begin{figure*}
\includegraphics[width=1\linewidth]{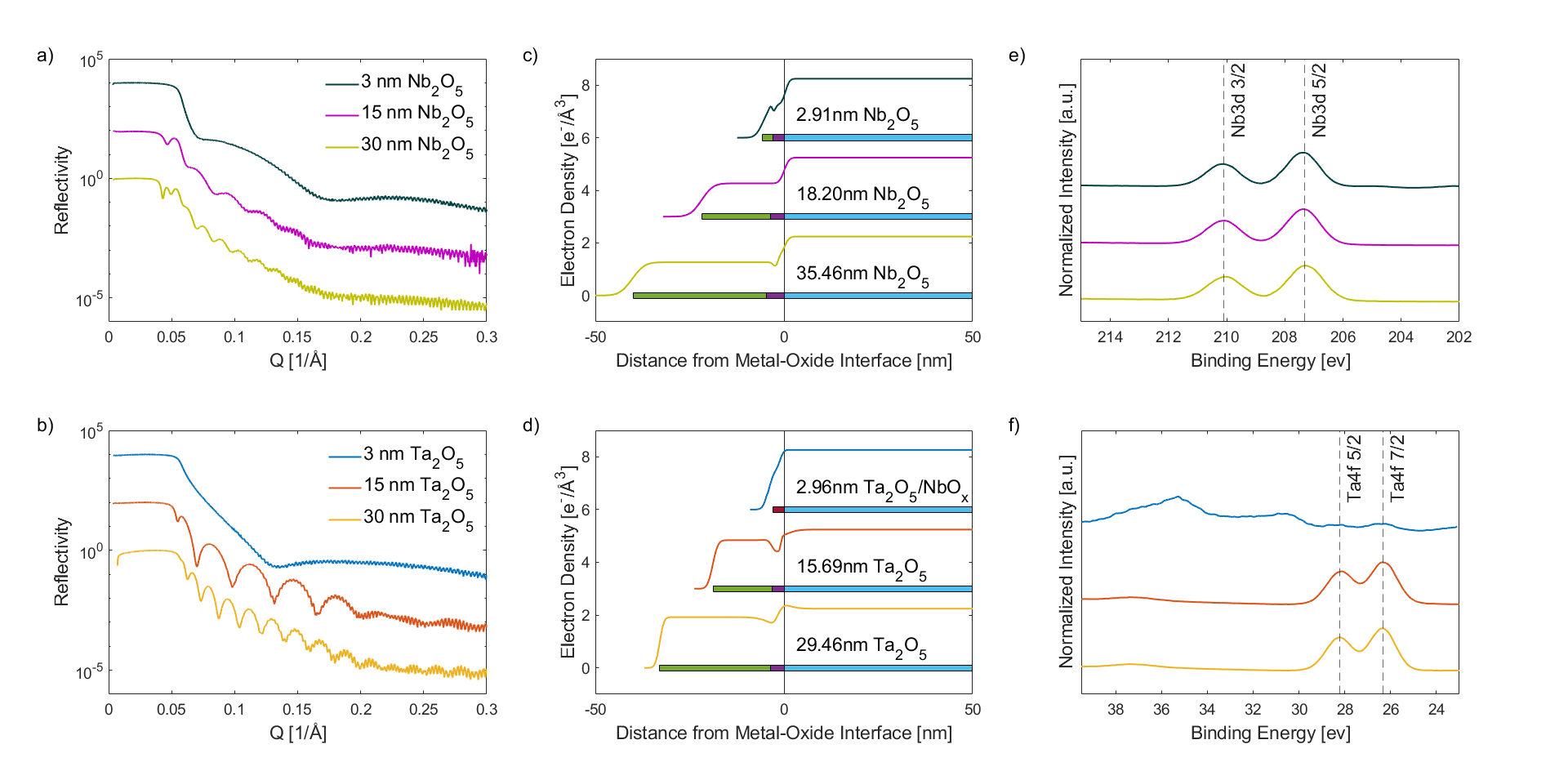}
\caption{\label{fig:XRR_eDensity_XPS}
Pentoxide X-ray reflectivity and photoelectron characterization. 
(a-b) X-ray reflectivity (XRR) patterns of resonator chips with (a) deposited $\mathrm{Nb_2O_5}$ and (b) deposited $\mathrm{Ta_2O_5}$. 
(c-d) Extracted electron density profiles, indicating the thickness of the deposited oxide (green), the native Nb oxide (purple), and, for the 3 nm $\mathrm{Ta_2O_5}$, the combined oxide layer (maroon), as measured by XRR. 
(e-f) X-ray photoelectron spectroscopy (XPS) of resonator chips with (e) deposited $\mathrm{Nb_2O_5}$ and (f) deposited $\mathrm{Ta_2O_5}$, measuring the Nb3d and Ta4f regions, respectively. Only peaks corresponding to the 5+ oxidation state of Nb and Ta are observed for the respective deposited oxides. For the 3 nm $\mathrm{Ta_2O_5}$ sample, additional Nb4p peaks are observed, corresponding to the underlying Nb film of the resonator.
}
\end{figure*}

To confirm the thicknesses of the deposited oxides, we perform XRR (Fig.~\ref{fig:XRR_eDensity_XPS}(a-b)). These thicknesses are shown in the corresponding electron density profiles shown in Fig.~\ref{fig:XRR_eDensity_XPS}(c-d). From the XRR, we observe homogeneity of the electron density in the oxide layers for the nominal 3 nm, 15 nm, and 30 nm $\mathrm{Nb_2O_5}$ devices and the 15 nm and 30 nm $\mathrm{Ta_2O_5}$ devices. For the 3 nm  $\mathrm{Ta_2O_5}$ device, we calculate a gradient in the electron density from the deposited $\mathrm{Ta_2O_5}$ to the native $\mathrm{Nb_2O_5}$. The thicknesses and discrete nature of the deposited oxide layers were further confirmed by STEM, as shown in Fig.~\ref{fig:TEM_thickness_SI}.

We also perform X-ray photoelectron spectroscopy (XPS) on the resonator devices. For nominal 15 nm and 30 nm samples, we observe only $\mathrm{Nb^{5+}}$ and $\mathrm{Ta^{5+}}$ in the Nb3d and Ta4f spectra, respectively. For 3 nm samples, we observe additional Nb peaks from the underlying Nb native oxide and metal. These additional features include the lower binding energy peaks in the Nb3d spectra (Fig.~\ref{fig:XRR_eDensity_XPS}(e)) and the higher binding energy Nb4p peaks in the Ta4f spectra (Fig.~\ref{fig:XRR_eDensity_XPS}(f)). This is consistent with a conventional $\sim$10 nm information depth of XPS.

\begin{figure}
\includegraphics[width=1\linewidth]{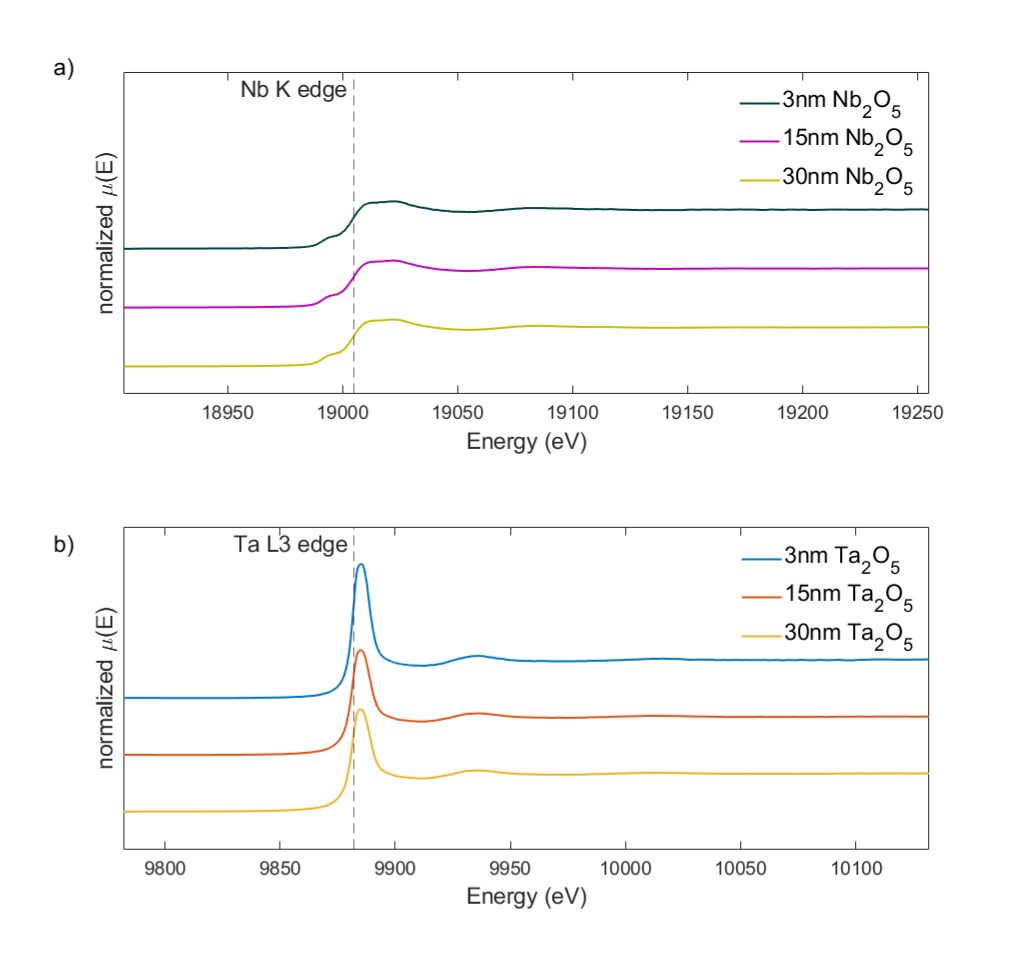}
\caption{\label{fig:XANES}X-ray absorption near-edge structure (XANES) spectra of the (a) Nb K edge for amorphous $\mathrm{Nb_2O_5}$ films and (b) Ta L3 edge for amorphous $\mathrm{Ta_2O_5}$ films, both deposited by pulsed laser deposition on Si(100) substrates. The position of the edge was determined to be the maximum in the first derivative of the spectrum. In both spectra, the position of the respective edge feature is constant, indicating a similar average chemical state of the Nb or Ta atoms, regardless of the thickness of the film.
}
\end{figure}

The chemical homogeneity throughout the pentoxide layer for all three thicknesses is further proven using X-ray absorption near-edge structure (XANES) spectroscopy on $\mathrm{Nb_2O_5}$ and $\mathrm{Ta_2O_5}$ thin films that were also deposited by PLD on Si (100) using identical deposition parameters as the oxides deposited on the resonator devices. These results are shown in Fig.~\ref{fig:XANES}. Unlike XPS, where the information depth is limited to the top few nanometers, XANES is not inherently surface sensitive and can provide information about the entire film.~\cite{Seah_2011, Iglesias-Juez_2021} For this reason, we choose to do XANES on the witness samples to isolate the signal of the deposited film from the underlying metal present in the resonator devices. From the XANES, we see similar pre-edge features, and identical shifts in the absorption edge, for the $\mathrm{Nb_2O_5}$ and $\mathrm{Ta_2O_5}$ samples, respectively. Since the spectra for the nominal 3 nm, 15 nm, and 30 nm samples have these similarities, this supports the homogeneity suggested by the EELS.

\section{Device measurement}

\begin{figure}
\includegraphics[width=\linewidth]{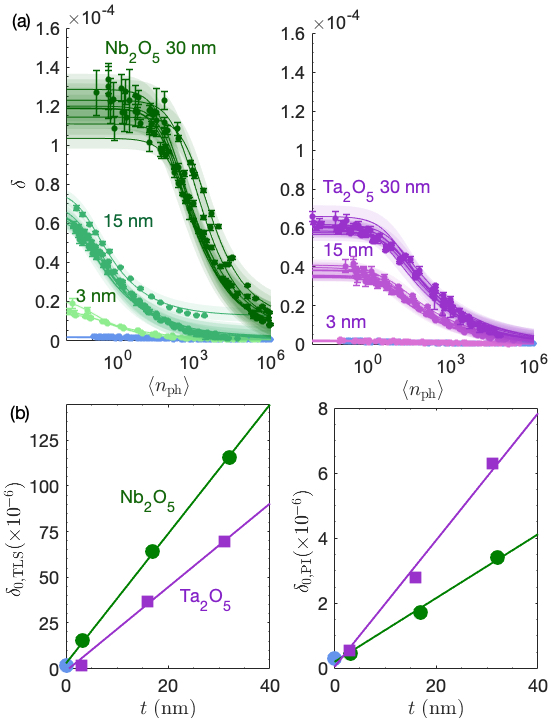}
\caption{\label{fig:device}Microwave measurements of pentoxide-coated 
superconducting resonators at a temperature of $\sim$10 mK. (a) Power-dependent loss tangent. Total resonator loss tangent $\delta$ as a function of average photon number $\langle n_{ph} \rangle$ in the resonator. Points are obtained by fitting complex $S_{21}$ data with the diameter correction method,~\cite{Khalil2012} with fit uncertainty shown as error bars. Lines represent fits to the resonator TLS loss tangent equation,~\cite{McRae2020} and 95$\%$ confidence prediction curves are shown as shaded regions. (b) Linear fits of TLS and PI loss tangents for Nb and Ta pentoxide-coated samples. Blue circle represents the Nb on sapphire control sample, green circles represent Nb pentoxide-coated samples, and purple squares represent Ta pentoxide-coated samples. Pentoxide top surface thickness $t$ for each sample is determined by STEM and shown in the supplementary material.
}
\end{figure}

Fig.~\ref{fig:device} shows the results of $S_{21}$ microwave measurements of seven chips with four to eight feedline-coupled $\lambda/4$ CPW resonators measured from each chip. VNA transmission measurements are performed around 10 mK in a dilution refrigerator with 60 dB of cryogenic attenuation. 

Measurements of the Nb control sample (no added oxide) show a loss tangent an order of magnitude lower than five of the six pentoxide-coated samples, allowing strong sensitivity to the pentoxide loss in this study.

Power-dependent measurements of the samples (Fig.~\ref{fig:device}(a)) definitively show that samples coated with Ta pentoxide are less lossy than those coated with a similar thickness of Nb pentoxide. 

To quantify the loss tangent introduced to the resonator per nm of pentoxide, linear fits are performed (Fig.~\ref{fig:device}(b)). The alignment of the y-intercept of these curves with the TLS loss tangent of the control sample $\delta_{ctrl}$ is evidence for the homogeneity of the dielectric loss tangent throughout the deposited oxide films: the pentoxide surface and interface with the underlying device is not contributing significantly here. From the linear fit, we see that depositing $\mathrm{Nb_2O_5}$ ($\mathrm{Ta_2O_5}$) adds TLS loss of $3.5$ $(2.3) \times 10^{-6}$ per nm of oxide, with $\mathrm{Nb_2O_5}$ adding about 30$\%$ more TLS loss than $\mathrm{Ta_2O_5}$. 

The thickness of the pentoxide on the top surface of the CPW conductor is used to perform these fits, which, as shown in Fig.~\ref{fig:TEM_thickness_SI}, is thicker than the tapered sidewall layer. We choose to use the top surface thickness here as it is easier to quantify. However, using the sidewall values, we arrive at the same relative loss difference between the Nb and Ta pentoxide films.

This result is consistent with the conclusions of  Ref.~\citenum{Pritchard_2025}, where it is demonstrated that there is a possible type of TLS that is present in $\mathrm{Nb_2O_5}$ and not $\mathrm{Ta_2O_5}$; namely, hyperfine couplings occurring between Nb nuclei and local magnetic moments in the pentoxide. 

We see a similar linear relationship in the power-independent (PI) losses, with $\mathrm{Nb_2O_5}$ ($\mathrm{Ta_2O_5}$) adding PI loss of $0.10$ $(0.20) \times 10^{-6}$ per nm of oxide. Here, while $\mathrm{Ta_2O_5}$ contributes half as much PI loss as $\mathrm{Nb_2O_5}$, both are negligible compared to the added TLS loss. The reason for this linear trend in PI loss is not obvious, but may represent a PI loss mechanism that is induced by TLS.

The nominal 3 nm $\mathrm{Ta_2O_5}$ sample may be seen as a single outlier point - it does not seem to have significantly more TLS loss than the control device. This may be explained in part by the gradient in the electron density seen in XRR in Fig.~\ref{fig:XRR_eDensity_XPS}(d). 

\section{Conclusion}

In this work, we deposit three thicknesses of amorphous $\mathrm{Nb_2O_5}$ and $\mathrm{Ta_2O_5}$ films onto test coplanar waveguide resonator devices. The pentoxide films are shown to be homogeneous and representative of natively forming $\mathrm{Nb_2O_5}$ and $\mathrm{Ta_2O_5}$. 

We measure the dielectric loss tangent of $\mathrm{Nb_2O_5}$ to be about 30$\%$ higher than that of $\mathrm{Ta_2O_5}$, demonstrating that the lack of suboxides is not the only difference between native Nb and Ta oxide. We hypothesize that the additional TLS loss in $\mathrm{Nb_2O_5}$ could be due to TLS caused by hyperfine splitting, as discussed in Ref.~\citenum{Pritchard_2025}.

\begin{acknowledgments}

This work was supported by the U.S. Department of Energy, Office of Science, National Quantum Information Science Research Centers, Superconducting Quantum Materials and Systems Center (SQMS), under Contract No. 89243024CSC000002.

This work made use of the Pulsed Laser Deposition Shared Facility at the Materials Research Center at Northwestern University, RRID:SCR\_017889, supported by the National Science Foundation MRSEC program (DMR-2308691) and the Soft and Hybrid Nanotechnology Experimental (SHyNE) Resource (NSF ECCS-2025633).

This work made use of the Northwestern University Jerome B. Cohen X-ray Diffraction Core Facility (RRID:SCR\_017866) supported by the MRSEC program of the National Science Foundation (DMR-2308691) at the Materials Research Center of Northwestern University and the Soft and Hybrid Nanotechnology Experimental (SHyNE) Resource (NSF ECCS-2025633).

This work made use of the EPIC (RRID: SCR\_026361), Keck-II (RRID: SCR\_026360), and NUFAB (RRID:SCR\_017779) facilities of Northwestern University’s NUANCE Center, which has received support from the IIN and Northwestern's MRSEC program (NSF DMR-2308691).

Part of this research used 6BM BMM beamline resources of the National Synchrotron Light Source II, a U.S. Department of Energy (DOE) Office of Science User Facility operated by Brookhaven National Laboratory under Contract No. DE-SC0012704.

This work was conducted with the support of funding through the National Institute of Standards and Technology (NIST). Certain commercial materials and equipment are identified in this paper to foster understanding. Such identification does not imply recommendation or endorsement by NIST, nor does it imply that the materials or equipment identified is necessarily the best available for the purpose.

We acknowledge support from IARPA and MIT Lincoln Laboratory for providing the TWPA used in this experiment.

Thanks to Eva Gurra and Tony McFadden for helpful feedback.

\end{acknowledgments}

\section*{Data Availability Statement}


The data that support the findings of this study are available from the corresponding author upon reasonable request.

\appendix

\section{Base CPW device fabrication details}

A substrate of HEMEX sapphire is sonicated in NMP at 80 $^\circ$C for 5 min, followed by 5 min in acetone and 5 min in IPA. Then, the substrate is submerged in a 5:1:1 DI:$\mathrm{NH_4OH}$:$\mathrm{H_2O_2}$ solution at 60 $^\circ$C for 3 min. The substrate is then loaded into the DC sputter system and outgassed for 1 h at a nominal temperature of 250 $^\circ$C. Temperature is then lowered to ambient, and 200 nm of Nb is sputter-deposited on the substrate surface.

Structures are patterned by exposing i-line photoresist with a maskless aligner (MLA) and developing with TMAH. Nb etching is performed with a fluorine-based reactive ion etch timed using laser endpoint detection. Wafers are exposed to a 2 min 60 W $\mathrm{O_2}$ ash before stripping and dicing.

\section{STEM characterization details}

\begin{figure}
\centering
\includegraphics[width=1\linewidth]{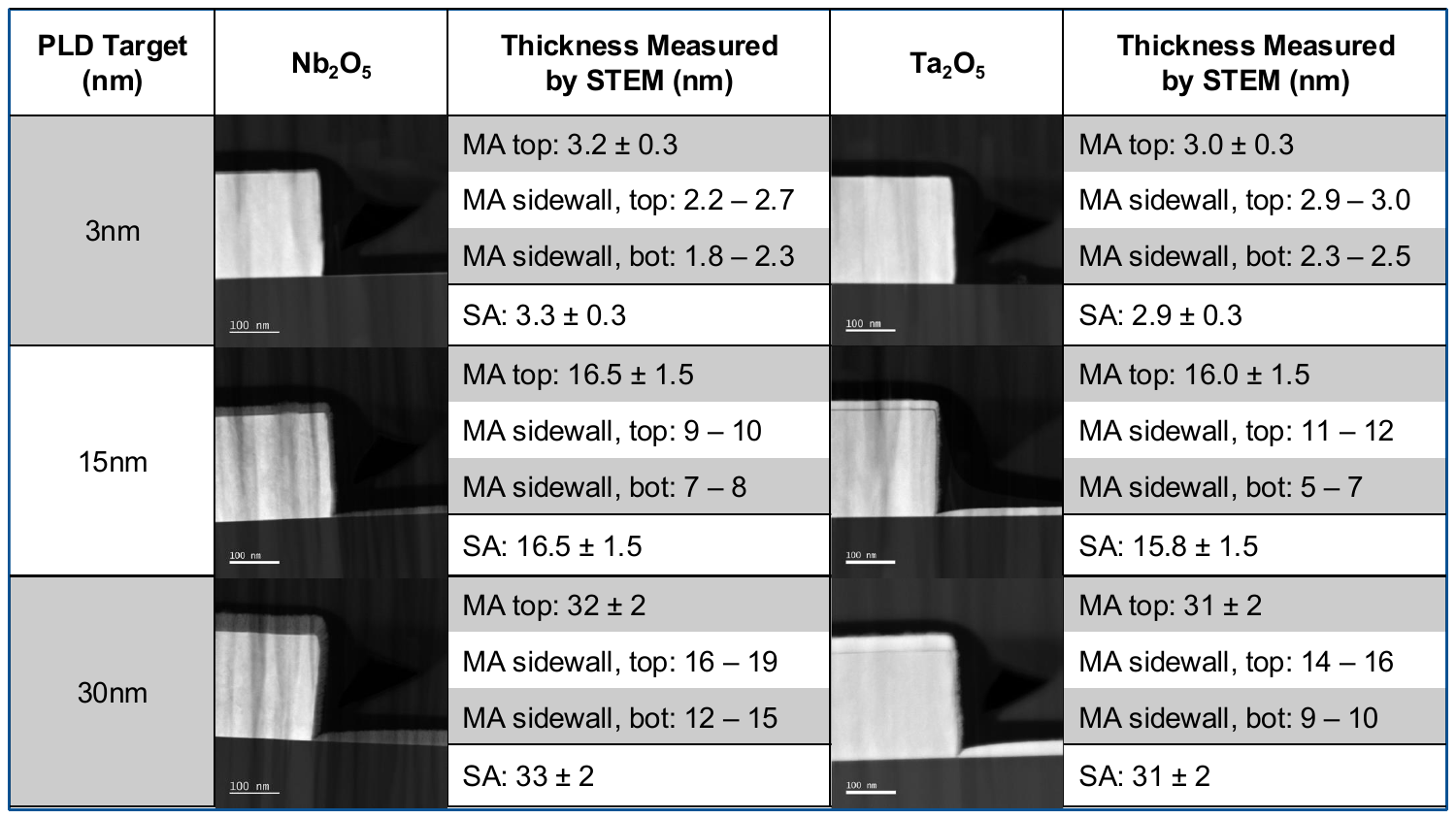}
\caption{\label{fig:TEM_thickness_SI}
HAADF-STEM images of resonator sidewalls for each of the six devices with nominally 3, 15, and 30 nm deposited $\mathrm{Nb_2O_5}$ and $\mathrm{Ta_2O_5}$, as described in the main text. The thicknesses of the deposited oxides measured from these images are additional corroboration of the thickness values extracted from XRR in Fig. \ref{fig:XRR_eDensity_XPS}, and serve as a more accurate representation of the overall volume of the pentoxides when measuring and extracting the loss tangents in Fig. \ref{fig:device}. Variations in the measured pentoxide thicknesses along the MA-top and SA interfaces are due to the roughness of the Nb film, whereas variations along the sidewall are representative of gradual changes in the actual deposited thickness due to self-shadowing effects in the PLD.
}
\end{figure}

Cross-sectional lamellas for devices with nominally 30 nm thick Nb and Ta pentoxides  (shown in Fig.~\ref{fig:matchar}) are prepared by focused ion beam (FIB) using the FEI Helios Nanolab 600 DualBeam FIB/SEM. To minimize charging and drifting effects in the FIB, 200 nm of Ni discharge layer is coated on the samples before FIB using the BOC Edwards Auto 500 electron-beam evaporator. Bulk-out, lift-out, and thinning processes are performed using a 30 kV Ga\textsuperscript{+} ion beam, while final cleaning steps are performed at 5 kV and 2 kV to remove amorphous materials on the surface. Additional cross-sectional lamellas for the second set of 6 devices (with nominally 30, 15, and 3 nm thick Nb and Ta pentoxides, shown in Fig.~\ref{fig:TEM_thickness_SI}) are also prepared by FIB using the Thermo Scientific Helios 5 Hydra CX DualBeam Plasma FIB/SEM. Bulk-out, lift-out, and thinning processes are performed using a 30 kV Xe\textsuperscript{+} ion beam, while final cleaning steps are performed at 8 kV and 5 kV to remove amorphous materials on the surface. Different FIB tools were used based on equipment availability. We do not expect this difference to impact the observed structure or chemistry, and direct comparisons are only made between samples that have gone through identical preparation procedures. All samples are then plasma cleaned using South Bay Technology PC-2000 Plasma Cleaner with Ar plasma at approximately 40 W RF power and 150 mTorr to further remove residual contamination. All lamella thicknesses are approximately 50 to 80 nm. 

Scanning transmission electron microscopy (STEM) characterization is performed on an aberration-corrected JEOL JEM-ARM200CF S/TEM operating at 200 kV. The STEM convergence angle and high-angle annular dark-field (HAADF) collection angles are 27 mrad and 90-370 mrad, respectively. Energy-dispersive X-ray spectroscopy (EDS) data are collected in the same microscope at 200 kV with Thermo Fisher Dual SDD detectors (1.7 steradians). Electron energy-loss spectroscopy (EELS) data are also collected in the same microscope at 200 kV, using the Gatan GIF Quantum filter with K2 Summit direct electron detector. The measurements were performed using a 5 mm aperture and a 100 eV energy slit. The dispersion of the EELS measurements was 0.25 ev/Ch, giving an energy resolution of approximately 0.8 eV, according to the full width at half maximum of the zero-loss peak. Data processing (background subtraction, denoising, signal mapping and analysis, and plural scattering removal) is performed using the Gatan Microscopy Suite (GMS) software. In particular, an inverse power-law function is used to model the pre-edge region of the O K-edge for background subtraction, then Fourier-ratio deconvolution is used for plural scattering removal. For the ELNES analysis shown in Fig.~\ref{fig:matchar}(h-i), 20 line scans, each with a 0.25 nm width in real space, are summed to give an integration width of 5 nm, whereas for the EELS line profile analysis shown in Fig.~\ref{fig:matchar}(j-k), four line scans are summed to give an integration width of 1 nm for each spectrum.

All six variations of our samples were imaged in the electron microscope, shown in Fig.~\ref{fig:TEM_thickness_SI}, to accurately visualize the oxide coating profile. This ensures conformal coating of the deposited oxides, in particular along the sidewalls and their corners, for which the XRR is not sensitive. We find that while the oxide thicknesses across the top surface of the film are consistent with our nominal target values, the sidewall is coated considerably less than that, with variations along the profile.

\begin{figure}
\centering
\includegraphics[width=\linewidth]{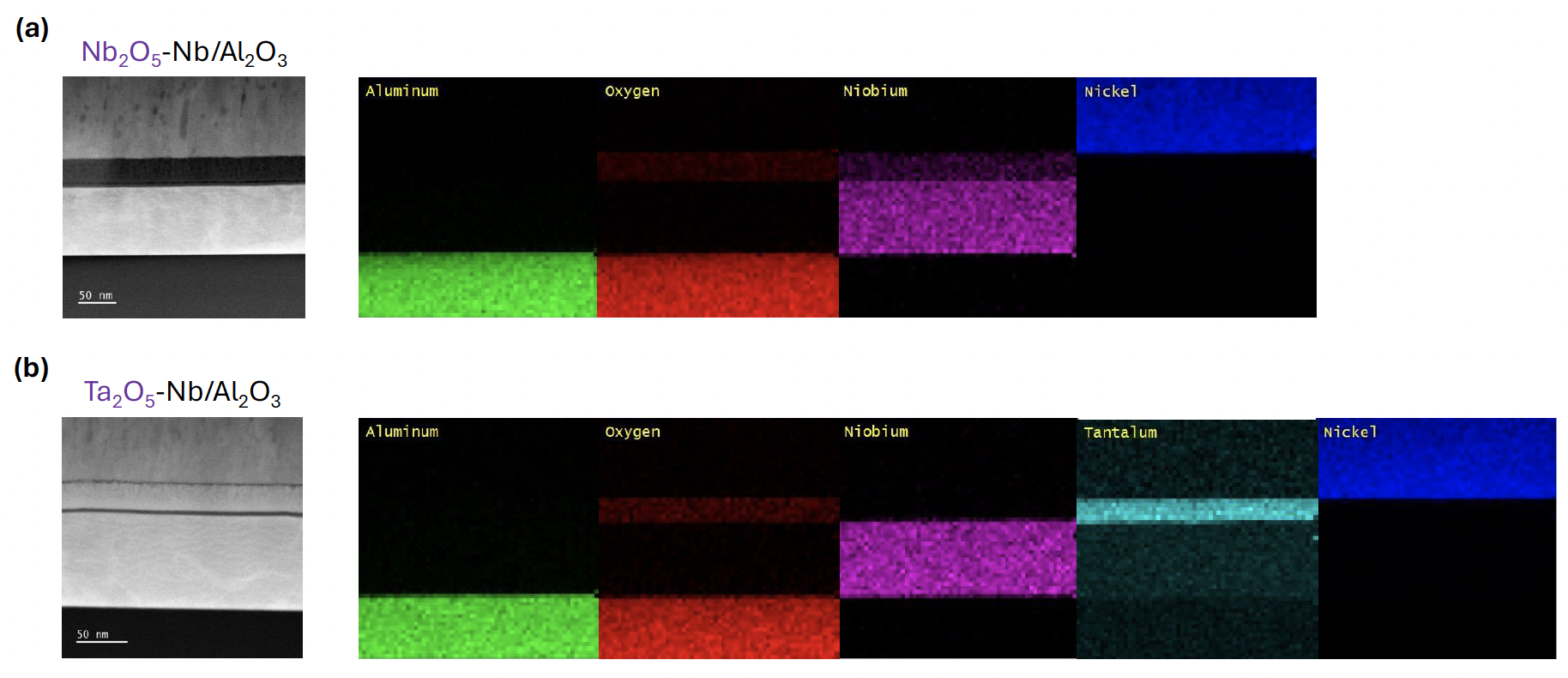}
\caption{\label{fig:TEM_EDS_SI}
EDS maps of the Nb films with 30 nm of deposited (a) $\mathrm{Nb_2O_5}$ and (b) $\mathrm{Ta_2O_5}$. These maps superimposed correspond to Fig. \ref{fig:matchar}(d) and \ref{fig:matchar}(e).}
\end{figure}

The elemental maps extracted from EDS, shown in Fig.~\ref{fig:TEM_EDS_SI} and superimposed in Fig.~\ref{fig:matchar}(d) and (e), confirm qualitatively that the chemical makeup of the layers is discrete; in other words, there is no widespread infiltration of oxygen from the pentoxides into the Nb bulk.

\section{Microwave measurement details}

The measurement wiring diagram for the FormFactor JDRY-250 used in this work is shown in Fig.~\ref{fig:bcqt_fridge}. Cryogenic switches are used to measure multiple devices using the same input and output lines. To improve SNR and measurement time at low power, a J-TWPA is included in the amplification chain.~\cite{Macklin2015} Samples in this work were measured over three successive cooldowns.

\begin{figure}
\label{fig:bcqt_fridge}
\includegraphics[width=\linewidth]{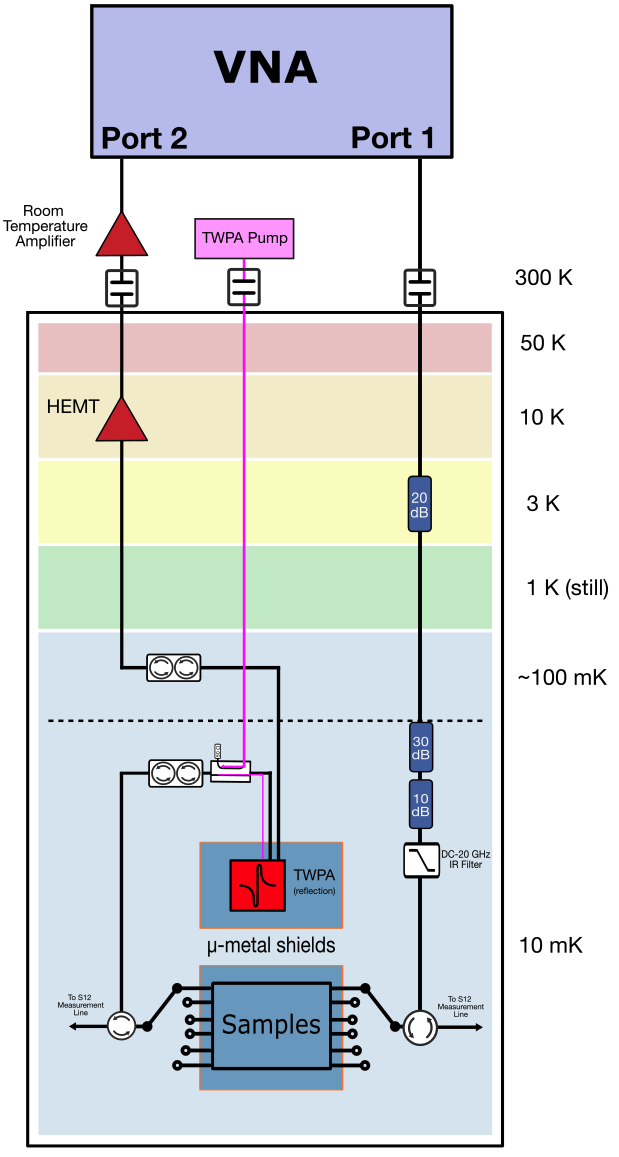}
\caption{\label{fig:bcqt_fridge} Wiring diagram for FormFactor JDRY-250 dilution refrigerator. Two input and two output lines are included, as noted by labels at the circulators.}
\end{figure}

\bibliography{pentoxides}

\end{document}